# On Distributions of Emission Sources and Speed of Sound in Proton-proton (Proton-antiproton) Collisions

Li-Na Gao and Fu-Hu Liu[1]

*Institute of Theoretical Physics, Shanxi University, Taiyuan, Shanxi 030006, China*

## Abstract

The revised (three-source) Landau hydrodynamic model is used in this paper to study the (pseudo)rapidity distributions of charged particles produced in proton-proton and proton-antiproton collisions at high energies. The central source is assumed to contribute with a Gaussian function which covers the rapidity distribution region as wide as possible. The target and projectile sources are assumed to emit isotropically particles in their respective rest frames. The model calculations obtained with a Monte Carlo method are fitted to the experimental data over an energy range from 0.2 to 13 TeV. The values of the squared speed-of-sound parameter in different collisions are then extracted from the width of the rapidity distributions.

## 1. Introduction

In hadron-hadron, hadron-nucleus, and nucleus-nucleus (heavy ion) collisions at high energies, the final-state particles could be produced by multiple emission sources (which we denote also fireballs) within the interacting system. This scenario can be tested by means of an analysis of the kinematic distributions of final-state particles. Proton-proton (proton-antiproton) collisions are usually used as a reference for the measurements in nucleus-nucleus collisions, in which the quark-gluon plasma (QGP) is expected to be formed [1]. From proton-proton collisions to nucleus-nucleus collisions, the distributions of emission sources may be similar to each other due to the small influences of nuclear effects such as the spectators and stopping power. The distributions of emission sources in high energy proton-proton (proton-antiproton) collisions can provide information on particle production in longitudinal rapidity and transverse momentum spaces.

In the study of distributions of emission sources, a principal question is whether there is a central source at midrapidity. If yes, what is the dependence of the relative contribution to particle production of the central source on collision energy? If no, what is the dependence of the width of the gap between two neighbouring sources around midrapidity on collision energy? In experiments at available accelerators and colliders, do the distributions of emission sources change with energy? Generally, common phenomenological models are useful to answer these questions. We focus on a few phenomenological models such as the three-fireball model [2-7], the three-source relativistic diffusion model [8-11], the multisource thermal model [12-14], the model with two Tsallis (or Boltzmann-Gibbs) clusters of fireballs [15-17], and the Landau hydrodynamic model [18-21] which results in a Gaussian shape for the rapidity distribution [21, 22] and a few revised versions such as works of Gao et al. [23], Wong [24], Jiang et al. [25-30], Beuf et al. [31], and Bialas et al. [32].

---

The three-fireball model [2-7] assumes the nucleon to be an extended object and to consist of valence quarks, sea quarks, and gluons. The interacting system formed in high energy proton-proton collisions can be divided into three fireballs: a central fireball located at midrapidity, a target fireball located in the backward (target) rapidity region, and a projectile fireball located in the forward (projectile) rapidity region. The three-source relativistic diffusion model [8-11] was also proposed to have three fireballs or sources: a central source located at midrapidity and arising from interactions between low-momentum gluons in both target and projectile, a target-like source located in the backward rapidity region and arising from interactions between valence quarks in the target and low-momentum gluons in the projectile, and a projectile-like source located in the forward rapidity region and arising from interactions between low-momentum gluons in the target and valence quarks in the projectile.

The multisource thermal model [12-14] is the successor of the thermalized cylinder model and the two-cylinder model. The thermalized cylinder model was proposed to have a (central) thermalized cylinder at midrapidity and two leading nucleon sources in target and projectile rapidity regions respectively. The two-cylinder model uses two (target/projectile) cylinders and two leading (target/projectile) nucleon sources. In the case of the two cylinders having a gap between them, there is no source at midrapidity. In most cases, there are sources at midrapidity due to the two cylinders overlapping each other or having no gap between them. In the model with two Tsallis clusters of fireballs [15], the rapidity distributions are described by using a superposition of two Tsallis fireballs along the rapidity axis. In the energy range from 0.5 to 7 TeV, the model result [15] shows that there is a gap between the two clusters, and there is no source at midrapidity. The Tsallis clusters can be replaced by others such as the Boltzmann-Gibbs clusters [16, 17].

The Landau hydrodynamic model [18-21] uses only a central source at midrapidity. The rapidity distribution obtained in the model is simply a Gaussian function [21, 22] in which the width is related to the speed of sound. A few revised versions were proposed in literature [22-32] to give better descriptions for experimental data, some of which include the contributions of leading and non-leading nucleons. We focus on a simple revised version [23] in which a central source at midrapidity, a target source in the backward rapidity region and a projectile source in the forward rapidity region are used. In the simple revised version, the central source is assumed to contribute with a Gaussian function which covers the rapidity distribution region as wide as possible. The target and projectile sources are assumed to emit isotropically particles in their respective rest frames.

From the above introduction, we see that most models were proposed to have a central source at midrapidity. In particular, under the assumption that the emitting sources are thermalized fireballs, the width of the rapidity distribution obtained by the (revised) Landau hydrodynamic model can be used to extract the squared speed-of-sound parameter, which is related to the mean free path of the strongly-interacting particles in the sources in the central, forward, and backward rapidity regions. We use the simple revised Landau hydrodynamic model [23] to describe the (pseudo)rapidity distribution. Because the widths of rapidity distributions obtained from the central source and from the target (or projectile) source are different, we expect to extract different values of the squared speed-of-sound parameter for the central and target (or projectile) sources.

In view of the wider application of the Landau hydrodynamic model [18-22] and its various revisions [22, 24-32], we have used the simple revised (three-source) Landau hydrodynamic model to study the pseudorapidity and rapidity distributions of charged particles produced in symmetric and asymmetric nuclear collisions at high energies in our recent work [23]. As a follow-up article of ref. [23], this work focuses on proton-proton (and proton-antiproton) collisions and presents a more detailed description of the model and its implementation. The definition of kinetic quantities is the same as in ref. [23] according to universal representations, in order to provide a consistent picture.

In this paper, we use the simple revised (three-source) Landau hydrodynamic model to study the pseudorapidity and rapidity distributions of charged particles produced in proton-proton (proton-antiproton) collisions at high energies. In section 2, a description of the model and calculation method is presented. Both pseudorapidity and rapidity distributions are obtained separately [33]. In section 3, the pseudorapidity distributions are compared with the experimental data of proton-proton (proton-antiproton) collisions over an energy range from 0.2 to 13 TeV [34-42]. The values of the squared speed-of-sound parameter are then extracted from the width of the rapidity distributions. In section 4, we summarize our main observations and conclusions.

## 2. The revised Landau hydrodynamic model and calculation method

The first 1+1-dimensional hydrodynamic model was proposed by Landau many years ago [18]. Later, a complex analytical solution was obtained by Khalatnikov [19]. The rapidity distribution of charged particles obtained from the complex analytical solution by Belenkij and Landau is [20]

$$\frac{dN_{ch}}{dy} \propto \exp\left(\sqrt{L^2 - y^2}\right), \tag{1}$$

where $L = \ln\left(\sqrt{s_{NN}}/2m_p\right)$ is the logarithmic Lorentz contraction factor, $\sqrt{s_{NN}}$ denotes the center-of-mass energy per pair of nucleons, and $m_p$ denotes the rest mass of a proton. A later study [21] showed that the rapidity distribution of charged particles follows a Gaussian function

$$\frac{dN_{ch}}{dy} \propto \exp\left(-\frac{y^2}{2L}\right) \tag{2}$$

in the case of $L \gg |y|$.

The second 1+1-dimensional hydrodynamic model and its analytical solution were obtained by Hwa [43] in the limit of $\sqrt{s_{NN}} \to \infty$. A plateau structure in rapidity distribution was obtained, which departs from available experimental results. Based on Hwa's work, Bjorken obtained the energy density of particles in high energy collisions [44]. By taking into account the contributions of leading particles and based on a theory of unified description of Hwa-Bjorken and Landau relativistic hydrodynamics, pseudorapidity distributions of charged particles in agreement with the experimental results were obtained in a recent work [23]. For the central source, the unified hydrodynamic model describes the rapidity distribution of charged particles to be [22]

$$\frac{dN_{ch}}{dy} = C \frac{|y|}{\left[\left(\theta_{FO} - \sqrt{\theta_{FO}^2 - c_s^2 y^2}\right)\left(\theta_{FO}^2 - c_s^2 y^2\right)\right]^{1/2}} \exp\left[-\frac{1-c_s^2}{2c_s^2}\left(\theta_{FO} - \sqrt{\theta_{FO}^2 - c_s^2 y^2}\right)\right], \tag{3}$$

where $C$ is the normalization constant, $c_s^2$ is the squared speed-of-sound, $\theta_{FO} = \ln(T_0/T_k)$, $T_0$ denotes the initial temperature, and $T_k$ denotes the kinetic freeze-out temperature.

Eq. (3) is similar to a Gaussian distribution [22]

$$\frac{dN_{ch}}{dy} = \frac{N_0}{\sqrt{2\pi}\sigma} \exp\left[-\frac{(y-y_C)^2}{2\sigma^2}\right], \tag{4}$$

where $N_0$ is the normalization constant, $\sigma$ is the distribution width, and $y_C$ is the midrapidity. In symmetric collisions, $y_C = 0$ in the center-of-mass reference frame. In our simple revised version of the Landau hydrodynamic model [23], $\sigma$ should be large enough to cover the rapidity region as wide as possible. The relation between $c_s^2$ and $\sigma$ can be given by [18, 24, 45-49]

$$\sigma = \sqrt{\frac{8}{3} \frac{c_s^2}{1 - c_s^4}} L \; . \tag{5}$$

Then, $c_s^2$ is expressed by using $\sigma$ to be

$$c_s^2 = \frac{1}{3\sigma^2} \left( \sqrt{16L^2 + 9\sigma^4} - 4L \right). \tag{6}$$

To perform the calculation for (pseudo)rapidity distribution as accurately as possible, we need also the transverse momentum ($p_T$) distribution. Here we use the simplest Boltzmann distribution [50]

$$f_{p_T}(p_T) = C_0 p_T \exp\left( -\frac{\sqrt{p_T^2 + m_0^2}}{kT_B} \right), \tag{7}$$

where $C_0$ is the normalization constant, $k$ denotes the Boltzmann constant, $T_B$ is the effective temperature which is larger than the kinetic freeze-out temperature $T_k$, and $m_0$ denotes the rest mass of the considered particle. The chemical potential and the distinction for fermions and bosons are not included due to small effects on transverse momentum distribution at high energy. Eq. (7) means that we assume thermal emission of the final state particles.

We introduce the transverse momentum distribution in the model so that we can obtain rapidity and pseudorapidity separately. In fact, to convert between rapidity and pseudorapidity distributions, additional limit is needed. The Boltzmann distribution describes the most fraction of the transverse momentum distribution. Eq. (7) is not the sole choice for the transverse momentum distribution. In fact, we can also use other choices such as the Tsallis distribution [51, 52], the Tsallis form of standard distribution [53, 54], the Erlang distribution [55], and so forth. In the study of rapidity or pseudorapidity distribution, the form of Eq. (7) and the value of parameter in it are not sensitive factors. Instead, the distributions of emission sources influence largely the rapidity or pseudorapidity distribution.

In the simple revised version of the Landau hydrodynamic model, we use in fact three sources: a central (C) source described by Eqs. (4) and (7), a target (T) source described by an isotropic emission picture and Eq. (7), and a projectile (P) source equals to the target source. The central source describes the contributions of all produced particles, partly non-leading nucleons, and all leading nucleons, while the target and projectile sources describe the contributions of partly non-leading nucleons. In the rapidity space, the central source stays at midrapidity $y_C$ and the target (projectile) source stays at rapidity $y_T$ ($y_P$) in the backward (forward) target (projectile) region.

We used a Monte Carlo method to perform the calculation for the central source. Let $R_{1,2,3}$ denote random numbers in $[0,1]$. According to Eqs. (4) and (7), we have $y = \sigma\sqrt{-2\ln R_1} \cos(2\pi R_2) + y_C$ and $\int_0^{p_T} f_{p_T}(p_T) dp_T < R_3 < \int_0^{p_T + dp_T} f_{p_T}(p_T) dp_T$ respectively. The longitudinal momentum is $p_z = \sqrt{p_T^2 + m_0^2} \sinh y$, the momentum $p = \sqrt{p_T^2 + p_z^2}$, and the pseudorapidity

$$\eta = \frac{1}{2} \ln\left( \frac{p + p_z}{p - p_z} \right). \tag{8}$$

In the calculation for the target and projectile sources, particles are assumed to be emitted isotropically in their respective rest frames. In the Monte Carlo method, the emission angle

$$\theta' = \arctan\left[ \frac{2\sqrt{R_4(1 - R_4)}}{1 - 2R_4} \right] + \theta_0, \tag{9}$$

where $\theta_0 = 0$ (or $\pi$) in the case of the first term in the above equation being larger than 0 (or smaller than 0) and $R_4$ denotes a random number in [0,1]. This isotropic emission results approximately in a Gaussian pseudorapidity distribution with the width of 0.91-0.92 [56], that can be compared with the theory of unified description of Hwa-Bjorken and Landau relativistic hydrodynamics [22], in which the rapidity of particle is assumed to obey a Gaussian distribution with the width of 0.85. In the present work, we do not need to study further the width corresponding to the target and projectile sources due to the fixed value of 0.91.

The transverse momentum $p_T$ of the particle produced in the target or projectile source has the same expression as those in the central source. The longitudinal momentum $p_z' = p_T \cot\theta'$, the momentum $p' = \sqrt{p_T^2 + p_z'^2}$, and the energy $E' = \sqrt{p'^2 + m_0^2}$ in the rest frame of the considered source can be extracted. In the laboratory or center-of-mass reference frame, the rapidity $y$ is given by

$$y = \frac{1}{2}\ln\left(\frac{E' + p_z'}{E' - p_z'}\right) + y_{T,P} \,. \tag{10}$$

The longitudinal momentum $p_z$, the momentum $p$, and the pseudorapidity $\eta$ have the same expressions as those for the central source.

In the above discussions, the rapidity distribution, $dN_{ch}/dy$, and the pseudorapidity distribution, $dN_{ch}/d\eta$, can be obtained separately, where $N_{ch}$ denotes the multiplicity of charged particles.

## 3. Comparisons with experimental data and discussion

The pseudorapidity distributions, $dN_{ch}/d\eta$, of charged particles produced in proton-proton ($p$-$p$) collisions at $\sqrt{s_{NN}} = 0.2$, 0.41, 0.9, 2.36, 7, and 13 TeV are presented in Figure 1, where $\sqrt{s_{NN}}$ is simplified to $\sqrt{s}$ for proton-proton and proton-antiproton collisions. The squares represent the experimental data measured by the PHOBOS [34] (a,b), ALICE [35, 36] (c,d,g), and CMS [37, 38] (e,f,h) Collaborations, whereas the curves represent our model results calculated by means of the Monte Carlo method. The types of collisions, inelastic interactions (INEL) or non-single diffractive interactions (NSD), are marked in the panels. We take $m_0 = 0.174$ GeV/$c^2$ for the central source and $m_0 = 0.938$ GeV/$c^2$ for the target and projectile sources respectively. The former is estimated from an average weighted the masses and yields of $\pi^\pm$, $K^\pm$, $\bar{p}$, and $p$ [57]; the latter is the mass of a proton due to protons being the charged particles in the target (projectile) source.

The effective temperature $T_B$ in Eq. (7) is taken to be 0.156 GeV [58, 59], which is the value extracted from a statistical hadronization model (a Grand-Canonical thermal model) analysis of lead-lead collisions at 2.76 TeV. This value is the chemical freeze-out temperature which can be extended to nucleus-nucleus collisions over an energy range from the critical energy (the minimum center-of-mass energy required to reach the phase transition to the QGP, which is thought to be in the range 10-20 GeV [58, 59]) to TeV and shows decrease for decreasing $\sqrt{s_{NN}}$. The energy range (0.2−13 TeV) considered in the present work is above the critical energy. As a small system, the collision discussed in the present work should have a chemical freeze-out temperature $T_{ch}$ which is less than 0.156 GeV. As the kinetic freeze-out temperature, $T_k$ should be lower than $T_{ch}$, due to final-state elastic rescattering of the hadrons. And as the effective temperature in Eq. (7), $T_B$ is larger than $T_k$, due to the effect of radial flow. It is hard to say the relative size between $T_B$ and 0.156 GeV, due to $T_B > T_k < T_{ch} < 0.156$ GeV. We assume $T_B$ to be 0.156 GeV and independent of energy in 0.2−13 TeV. In the fits for the rapidity and pseudorapidity distributions, $T_B$ is not a very sensitive quantity, and a relative large (~10%) increase or decrease in $T_B$ does not cause a very obvious change in the distributions. As an example, the dotted and dashed curves in Figure 1(h) are the fits with $1.1T_B$ ($\approx 0.172$ GeV) and $0.9T_B$ ($\approx 0.140$ GeV) respectively.

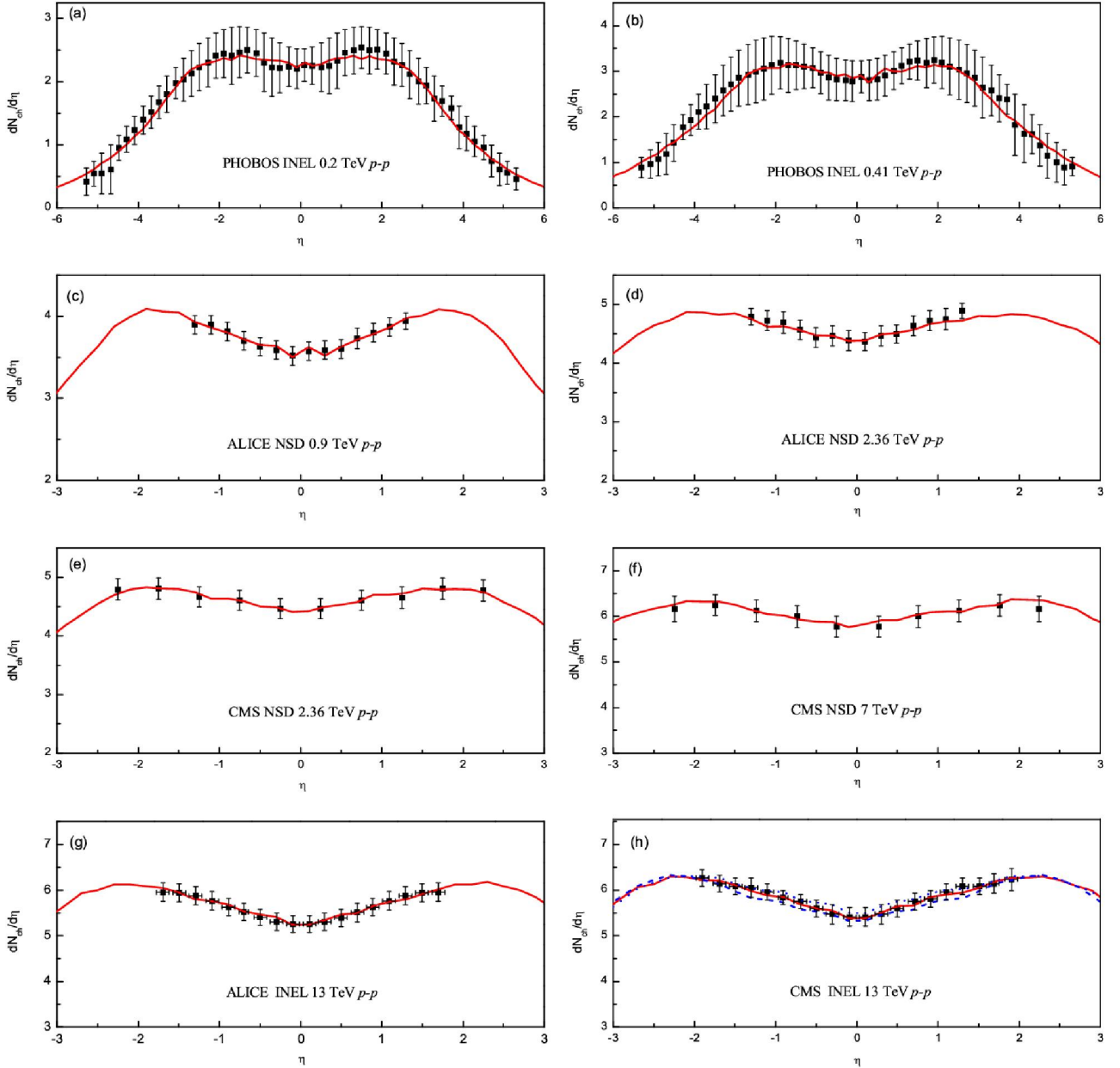

Figure 1. Pseudorapidity distributions of charged particles produced in *p-p* collisions at high energies. The squares represent the experimental data of the PHOBOS [34] (a,b), ALICE [35, 36] (c,d,g), and CMS [37, 38] (e,f,h) Collaborations, whereas the curves represent our fitted results.

The values of the peak position $y_T$ and relative contribution $k_T$ of the target source, the rapidity distribution width $\sigma$ of the central source, the normalization constant $N_0$, and $\chi^2$ per degree of freedom ($\chi^2/dof$) fitted by the method of least squares are given in Table 1, together with the squared speed-of-sound $c_s^2(T)$ for the target source and the squared speed-of-sound $c_s^2(C)$ for the central source. For *p-p* collision, the peak position $y_P$ and relative contribution $k_P$ of the projectile source, the peak position $y_C$ and the relative contribution $k_C$ of the central source, and the squared speed-of-sound $c_s^2(P)$ for the projectile source are given by $y_P = -y_T$, $k_P = k_T$, $y_C = 0$, $k_C = 1 - k_T - k_P$, and $c_s^2(P) = c_s^2(T)$, respectively.

Table 1. Fitted values of $y_T$, $k_T$, $\sigma$, $N_0$, and $\chi^2/dof$ corresponding to the curves in Figures 1 and 2 which show the pseudorapidity distributions in $p$-$p$ and $p$-$\bar{p}$ collisions at different energies. The last two columns show $c_s^2(T)$ and $c_s^2(C)$.

| Figure | $\sqrt{s}$ (TeV) | $y_T$ | $k_T$ | $\sigma$ | $N_0$ | $\chi^2/dof$ | $c_s^2(T)$ | $c_s^2(C)$ |
|---|---|---|---|---|---|---|---|---|
| Figure 1(a) | 0.2 | $-1.85\pm0.17$ | $0.027\pm0.012$ | $2.75\pm0.02$ | $20.50\pm0.37$ | 0.207 | $0.066\pm0.010$ | $0.472\pm0.050$ |
| Figure 1(b) | 0.41 | $-1.73\pm0.12$ | $0.020\pm0.008$ | $3.15\pm0.07$ | $29.20\pm0.43$ | 0.221 | $0.057\pm0.009$ | $0.511\pm0.050$ |
| Figure 1(c) | 0.9 | $-1.23\pm0.12$ | $0.027\pm0.008$ | $3.03\pm0.02$ | $35.00\pm0.68$ | 0.007 | $0.050\pm0.008$ | $0.446\pm0.050$ |
| Figure 1(d) | 2.36 | $-1.72\pm0.05$ | $0.016\pm0.007$ | $3.37\pm0.03$ | $47.80\pm0.52$ | 0.013 | $0.043\pm0.006$ | $0.467\pm0.016$ |
| Figure 1(e) | 2.36 | $-1.71\pm0.06$ | $0.013\pm0.010$ | $3.32\pm0.04$ | $47.00\pm0.50$ | 0.003 | $0.043\pm0.006$ | $0.458\pm0.020$ |
| Figure 1(f) | 7 | $-2.16\pm0.13$ | $0.020\pm0.008$ | $3.50\pm0.06$ | $65.80\pm0.83$ | 0.009 | $0.038\pm0.006$ | $0.447\pm0.010$ |
| Figure 1(g) | 13 | $-1.76\pm0.02$ | $0.026\pm0.006$ | $3.63\pm0.06$ | $63.20\pm0.42$ | 0.008 | $0.035\pm0.001$ | $0.447\pm0.010$ |
| Figure 1(h) | 13 | $-1.78\pm0.03$ | $0.024\pm0.005$ | $3.66\pm0.04$ | $65.30\pm0.38$ | 0.011 | $0.035\pm0.001$ | $0.452\pm0.006$ |
| Figure 2(a) | 0.54 | $-1.32\pm0.17$ | $0.020\pm0.002$ | $2.44\pm0.02$ | $12.80\pm0.27$ | 0.070 | $0.055\pm0.001$ | $0.347\pm0.040$ |
| Figure 2(b) | 0.63 | $-1.70\pm0.05$ | $0.050\pm0.003$ | $3.05\pm0.01$ | $33.00\pm0.64$ | 0.466 | $0.053\pm0.001$ | $0.468\pm0.050$ |
| Figure 2(c) | 0.63 | $-2.70\pm0.24$ | $0.010\pm0.003$ | $3.30\pm0.09$ | $33.70\pm0.49$ | 0.192 | $0.053\pm0.001$ | $0.516\pm0.050$ |
| Figure 2(d) | 0.9 | $-0.85\pm0.03$ | $0.012\pm0.005$ | $3.08\pm0.04$ | $33.00\pm0.63$ | 0.028 | $0.050\pm0.001$ | $0.456\pm0.050$ |
| Figure 2(e) | 0.9 | $-0.80\pm0.08$ | $0.023\pm0.007$ | $3.12\pm0.03$ | $29.50\pm0.58$ | 0.070 | $0.050\pm0.001$ | $0.464\pm0.050$ |
| Figure 2(f) | 1.8 | $-1.75\pm0.08$ | $0.050\pm0.002$ | $3.10\pm0.03$ | $20.90\pm0.33$ | 0.784 | $0.045\pm0.001$ | $0.428\pm0.040$ |

Figure 2 is similar to Figure 1, but it shows the results in proton-antiproton ($p$-$\bar{p}$) collisions at $\sqrt{s} =$ 0.54, 0.63, 0.9, and 1.8 TeV. The solid squares and curves represent the experimental data measured by the UA1 [39] (a), CDF [40] (b,f), P238 [41] (c), and UA5 [42] (d,e) Collaborations and our model results respectively, and the open squares are reflected at $\eta = 0$. The obtained values of $y_T$, $k_T$, $\sigma$, $\chi^2/dof$, $c_s^2(T)$, and $c_s^2(C)$ are given in Table 1. The relations among different peak positions, relative contributions, and speeds of sound for $p$-$\bar{p}$ collision are the same as those for $p$-$p$ collision.

Figures 1 and 2 show that the model results describe the experimental data. In the considered energy range, (i) $y_T$ does not show an obvious dependence on the energy $\sqrt{s}$, or $y_T$ for $p$-$p$ ($p$-$\bar{p}$) collision has a slight decrease (increase) with increase of $\sqrt{s}$; (ii) $k_T$ does not show an obvious change with $\sqrt{s}$; (iii) $\sigma$ increases with increase of $\sqrt{s}$; (iv) $c_s^2(T) < c_s^2(C)$, and they being clearly flat as functions of $\sqrt{s}$.

To see clearly the dependences of $y_T$, $\sigma$, $c_s^2(T)$, and $c_s^2(C)$ on $\sqrt{s}$, we plot the parameter values listed in Table 1 in Figure 3. The symbols and lines represent the parameter values and linear fitting functions respectively. The intercepts, slopes, and $\chi^2/dof$ corresponding to the lines in Figure 3 are listed in Table 2. The conclusions obtained from Figure 1, Figure 2, and Table 1 can be also seen in Figure 3.

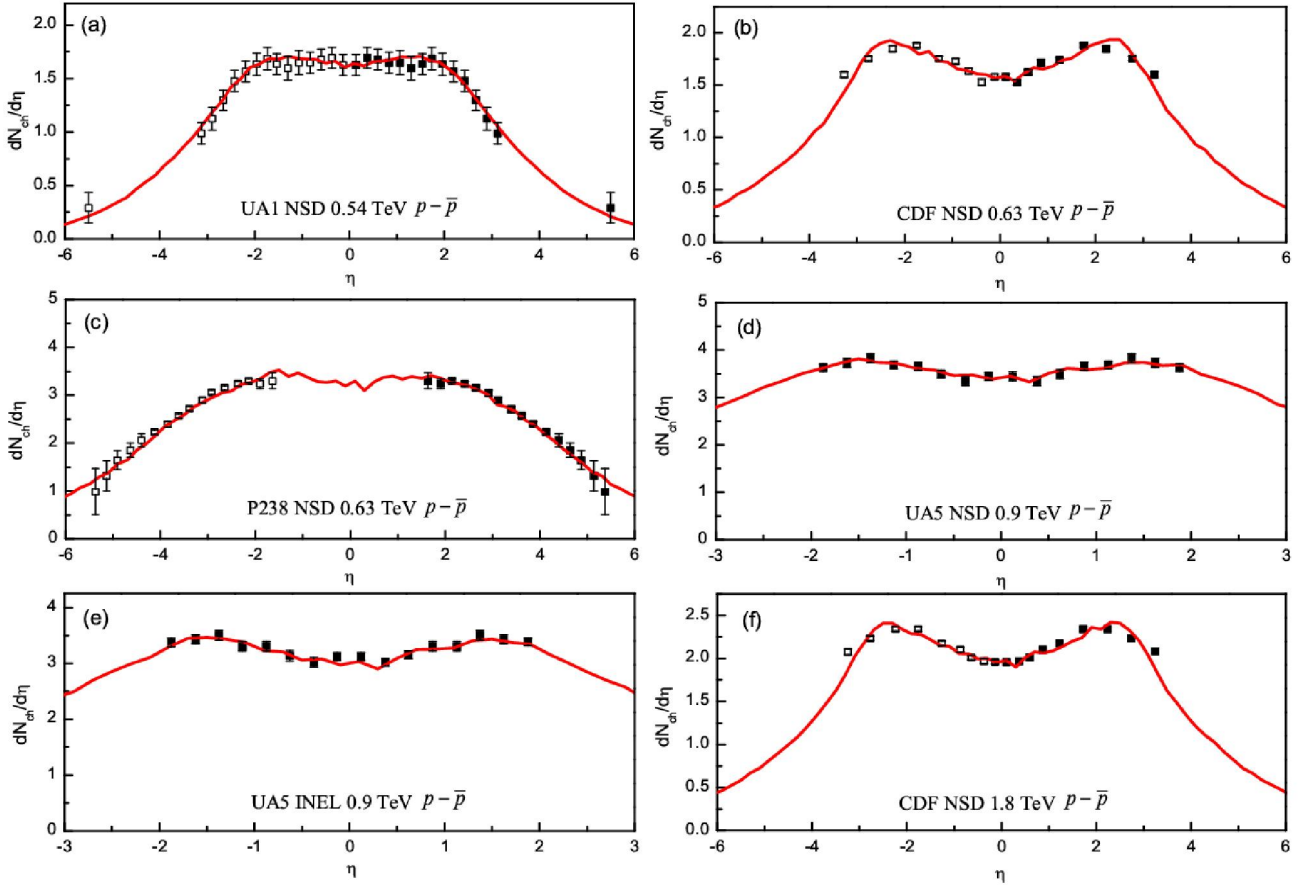

Figure 2. Pseudorapidity distributions of charged particles produced in $p$-$\bar{p}$ collisions at high energies. The solid squares and curves represent the experimental data of the UA1 [39] (a), CDF [40] (b,f), P238 [41] (c), and UA5 [42] (d,e) Collaborations and our fitted results respectively, and the open squares are reflected at $\eta = 0$ due to symmetry.

Table 2. Values of intercept, slope, and $\chi^2 / dof$ corresponding to the linear fits in Figure 3, though some of them do not obey linear relations. The unit of $\sqrt{s}$ is TeV.

| Figure | Type | Intercept | Slope | $\chi^2 / dof$ |
|---|---|---|---|---|
| Figure 3(a) | $y_T - \sqrt{s}$ | $-1.670 \pm 0.118$ | $-0.015 \pm 0.017$ | 9.697 |
| Figure 3(b) | $y_T - \sqrt{s}$ | $-1.593 \pm 0.673$ | $0.081 \pm 0.676$ | 157.282 |
| Figure 3(c) | $\sigma - \sqrt{s}$ | $3.061 \pm 0.080$ | $0.049 \pm 0.011$ | 55.317 |
| Figure 3(d) | $\sigma - \sqrt{s}$ | $2.852 \pm 0.270$ | $0.182 \pm 0.271$ | 188.858 |
| Figure 3(e) | $c_s^2(T) - \sqrt{s}$ | $0.054 \pm 0.003$ | $-0.002 \pm 0.001$ | 3.657 |
| | $c_s^2(C) - \sqrt{s}$ | $0.473 \pm 0.009$ | $-0.003 \pm 0.001$ | 1.159 |
| Figure 3(f) | $c_s^2(T) - \sqrt{s}$ | $0.058 \pm 0.001$ | $-0.007 \pm 0.001$ | 1.054 |
| | $c_s^2(C) - \sqrt{s}$ | $0.451 \pm 0.054$ | $-0.005 \pm 0.054$ | 2.179 |

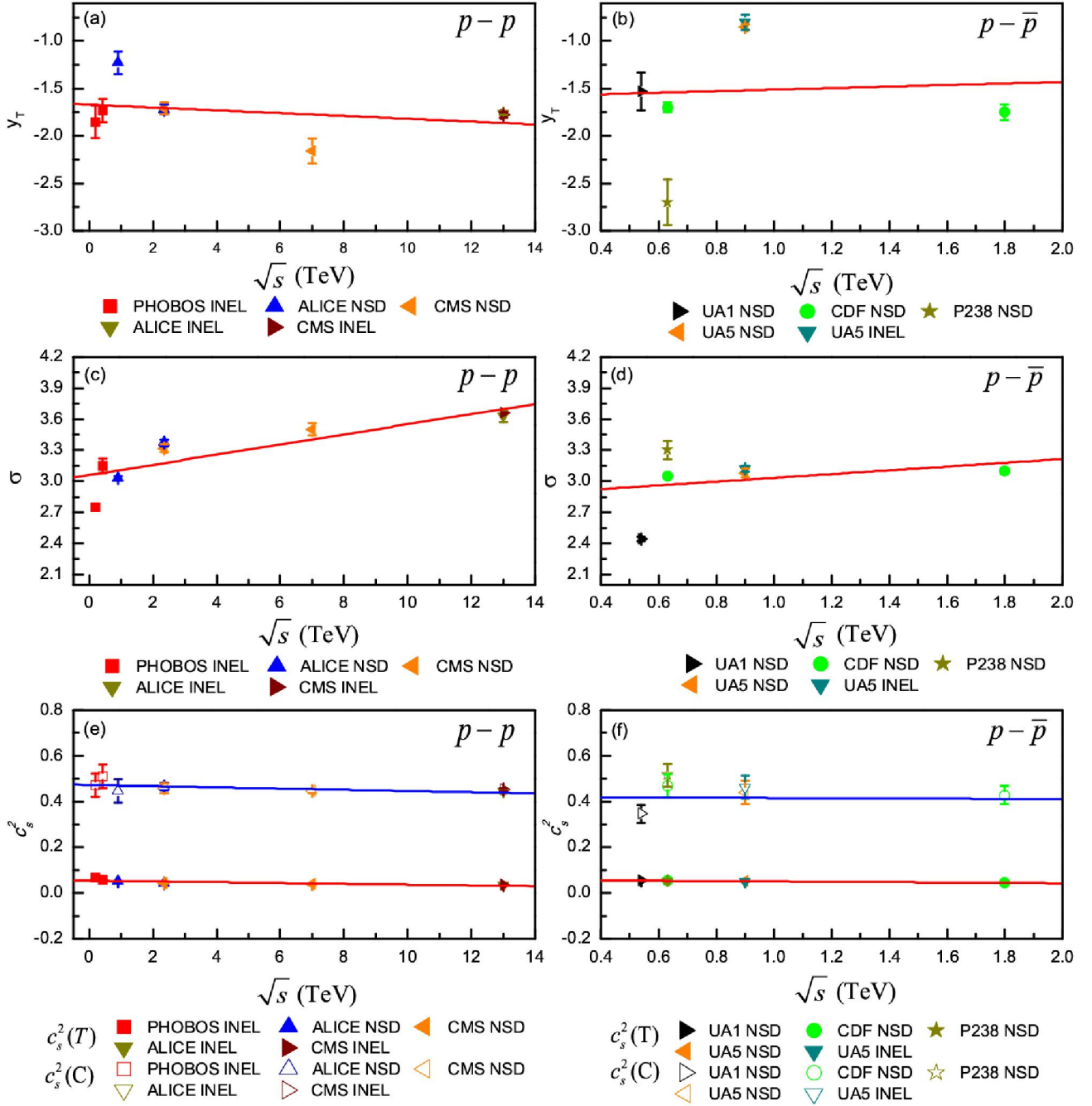

Figure 3. Collision energy dependence of $y_T$ (a,b), $\sigma$ (c,d), $c_s^2(T)$ (e,f), and $c_s^2(C)$ (e,f) for $p$-$p$ (a,c,e) and $p$-$\overline{p}$ (b,d,f) collisions. The symbols represent the values of free parameters and squared speed-of-sound listed in Table 1. The lines are our fitted results, though some of them do not obey linear relations.

In the above discussions, the physics picture and calculation method are the same as our previous work [23] which deals with symmetric and asymmetric nuclear collisions, and different from our other works [60, 61] in which four sources (target/projectile participant sources and target/projectile spectator sources) are used. In the present work, we have used a large enough central source which is described by the Landau hydrodynamic model [18-22] and two small target and projectile sources which revise the model. The rapidity range covered by the two participant sources is the same as that of the three fireballs [2-7] or three sources [8-11]; and the two spectator sources are beyond the scope of the compared picture, due to the former appearing in very backward or forward region. The rapidity range covered by the central source in the revised Landau hydrodynamic model is the same as that of the compared picture, and the target/projectile sources appear in the scope of the central source.

Because different methods are used, the present work obtains a different $c_s^2(C)$ ($\approx 1/3 \sim 1/2$) from our previous works [60, 61] ($\approx 0.15 \sim 0.28$). We observe that the new revised Landau hydrodynamic model yields larger values of $c_s^2(C)$. Let $D$ denote the dimensionality of space. We have the relation $c_s^2 = 1/D$ for massless particles [62, 63]. Then, the value $c_s^2 = 1/3$ for $D$=3 corresponds to the ideal gas state (large mean free path) of hadronic matter investigated by different groups in the past [64-68] and more recently [69-73]; instead, $c_s^2 = 1/2$ for $D$=2 corresponds to the ideal liquid state (small mean free path) investigated by Buchbinder et al. [63]. The present work seems to suggest that the target and projectile sources formed in proton-proton (proton-antiproton) collisions at $\sqrt{s} = 0.2 - 13$ TeV stay in the gas-like state and the central source stays in a liquid-like state. However, the QGP is an extended system in which colour confinement is removed (partonic degrees of freedom) and local thermal equilibrium is reached. In small collision systems, like proton-proton and proton-antiproton, the large values of $c_s^2(C)$ cannot be interpreted as a parameter of an extended thermalized system.

## 4. Conclusions

We summarize here our main observations and conclusions.

(a) As in nucleus-nucleus collisions, the interacting system formed in proton-proton (proton-antiproton) collision can be also divided into three emission sources. The central source is large enough to cover the whole (pseudo)rapidity region and is described by the Landau hydrodynamic model. The small target (projectile) source which is isotropic in its rest frame is used to revise the Landau hydrodynamic model and the (pseudo)rapidity distribution. Our treatment of the distributions of emission sources is a revision of the Landau hydrodynamic model. The model calculations obtained with a Monte Carlo method can describe the experimental data of proton-proton and proton-antiproton collisions over an energy range from 0.2 to 13 TeV.

(b) In proton-proton and proton-antiproton collisions in the considered energy range, the free parameters and the squared speed-of-sound $c_s^2$ do not show an obvious dependence on the energy. The values of $c_s^2(C)$ extracted for the central source are in the range from 1/3 to 1/2, and the values of $c_s^2(T)$ for the target and projectile sources are in the range from 0.04 to 0.07. It is interesting to note that the values of $c_s^2(C)$ in the range of 1/3 to 1/2 are expected to characterize the formation of a QGP in high energy collisions of heavy nuclei. The interpretation of large $c_s^2(C)$ in small systems such as in proton-proton and proton-antiproton collisions remains an open question that deserves further investigation.

## Conflict of Interests

The authors declare that there is no conflict of interests regarding the publication of this paper.

## Acknowledgments

The authors thank the reviewers and Editor for their comments and improvements to the manuscript. This work was supported by the National Natural Science Foundation of China under Grant no. 11575103, the Open Research Subject of the Chinese Academy of Sciences Large-Scale Scientific Facility under Grant no. 2060205, the Shanxi Provincial Natural Science Foundation under Grant No.

2013021006, and the Shanxi Scholarship Council of China under Grant no. 2012-012.

identified particles in Au+Au and d+Au collisions at $\sqrt{s_{NN}}$ =200 GeV," *Physical Review C*, vol. 88, Article ID 024906, 2013.